# DAΦNE STATUS AND UPGRADE PLANS

M. Zobov, INFN Laboratori Nazionali di Frascati, Frascati, Italy
for DAΦNE Collaboration Team[1]

*Abstract*

The Frascati Φ-factory DAΦNE has successfully completed experimental runs for the three main detectors, KLOE, FINUDA and DEAR. The best peak luminosity achieved so far is $1.6 \times 10^{32}$ cm$^{-2}$s$^{-1}$, while the best daily integrated luminosity is 10 pb$^{-1}$. At present the DAΦNE team is preparing an upgrade of the collider based on the novel crabbed waist collision scheme. The upgrade is aimed at pushing the luminosity towards $10^{33}$ cm$^{-2}$s$^{-1}$. In this paper we describe present collider performance and discuss ongoing preparatory work for the upgrade.

## INTRODUCTION

DAΦNE is an electron-positron collider working at the c.m. energy of the Φ resonance (1.02 GeV c.m.) to produce a high rate of K mesons [2]. The collider complex consists of two independent rings having two common Interaction Regions (IR) and an injection system composed of a full energy linear accelerator, a damping/accumulator ring and transfer lines. Figure 1 shows a view of the DAΦNE accelerator complex while some of the main collider parameters are listed in Table 1.

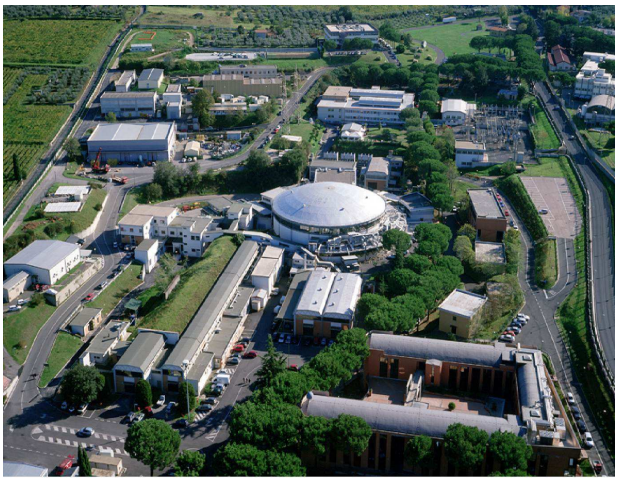

Fig. 1 View of DAΦNE accelerator complex.

Since year 2000 DAΦNE has been delivering luminosity to three experiments, KLOE [3], FINUDA [4] and DEAR [5]. The KLOE experimental detector surrounded by a superconducting solenoid has been used for a wide variety of physics measurements with emphasis on the kaon decays, and most notably on the issue of CP violation. The second magnetic detector FINUDA is devoted to the study of hypernuclear physics. The small non-magnetic experiment DEAR has been used for the study of the properties of kaonic atoms.

A wide spectrum of experiments is also carried out at the DAΦNE beam test facility (BTF) [6], a dedicated beam transfer line delivering electron or positron beams in the energy range 25-725 MeV with intensities varying from $10^{10}$ particle/pulse down to a single-electron. Besides, two separate beam lines are used for synchrotron radiation (SR) studies, extracting the SR light from a wiggler and a bending magnet, respectively [7].

Table 1: DAΦNE main parameters (KLOE run)

| | |
|---|---|
| Energy [GeV] | 0.51 |
| Trajectory length [m] | 97.69 |
| RF frequency [MHz] | 368.26 |
| Harmonic number | 120 |
| Damping time, $\tau_E/\tau_x$ [ms] | 17.8/36.0 |
| Bunch length [cm] | 1-3 |
| Number of colliding bunches | 111 |
| Beta functions $\beta_x/\beta_y$ at IP [m] | 1.6/0.017 |
| Emittance, $\varepsilon_x$ [mm·mrad] | 0.34 |
| Coupling [%] | 0.2-0.3 |
| Max. tune shifts | 0.03/0.04 |
| Max. beam current e-/e+ [A] | 2.4/1.4 |

At present DAΦNE is undergoing a shut down for installation of the SIDDHARTA experiment [8] that should start collecting data in 2008. The shut down foresees also modifications aimed at testing the novel idea of crab waist (CW) collisions [9].

## COLLIDER PERFORMANCE

Since the beginning of the experimental data taking runs in 2000 DAΦNE has been continuosly improving its performances in terms of luminosity, lifetime and backgrounds. Figure 2 shows the daily peak luminosity for KLOE, DEAR and FINUDA runs.

The DEAR experiment was performed in less than 5 months in 2002-2003, collecting about 100 pb$^{-1}$, with a peak luminosity of $0.7 \times 10^{32}$ cm$^{-2}$s$^{-1}$. The KLOE experimental program has been completed in year 2006, acquiring more than 2 fb$^{-1}$ on the peak of the Φ resonance, more than 0.25 fb$^{-1}$ off-resonance and performing a high statistics resonance scan. The best peak luminosity obtained during this run was $1.5 \times 10^{32}$ cm$^{-2}$s$^{-1}$, with a maximum daily integrated luminosity of about 10 pb$^{-1}$. The second run of FINUDA started in April 2006 and collected 0.96 fb$^{-1}$. During this run a peak luminosity of $1.6 \times 10^{32}$ cm$^{-2}$s$^{-1}$ has been achieved, while a maximum daily integrated luminosity similar to that in the KLOE run has been obtained with lower beam currents, lower number of bunches and higher beta functions at the collision point.

The steady luminosity progress shown in Fig. 2 was achieved by optimization of the machine parameters and hardware changes implemented during the long shut downs in 2003 and in 2006. The main accelerator physics issues that have been studied and optimized can be summarized as follows [10]:

- Working point choice.
- Coupling correction.
- Optimization of collision point parameters (betas, emittances, crossing angles etc.)
- Collider optics modelling.
- Nonlinear beam dynamics study.
- Single and multibunch instability cures.
- Fine tuning of all parameters in collisions.

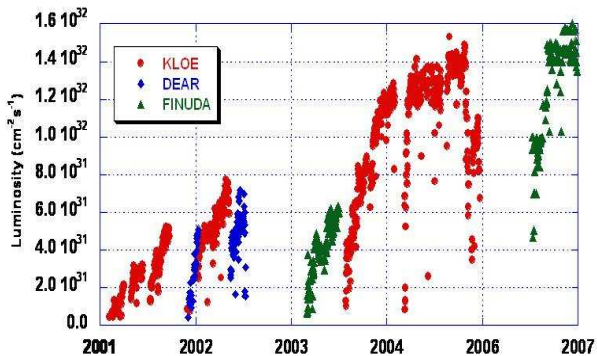

Figure 2. DAΦNE peak luminosity for KLOE (red), DEAR (blue) and FINUDA (green).

The most relevant hardware changes made during the 2003 shut down [11] were:

- New KLOE and FINUDA Interaction Regions.
- Long straight sections modification.
- Wiggler magnets modification.
- Shorter injection kicker pulse.
- Additional feedback amplifiers.
- Maintenance of bellows and scrapers.

The hardware changes and modifications implemented during the KLOE removal and FINUDA detector installation in 2006 [12] were:

- New simplified interaction region installation
- Wiggler ion clearing electrodes (ICE) removal
- New parasitic crossings (PC) compensating wires
- New BPMs for single turn measurements
- New 3$^{rd}$ generation transverse feedbacks
- Control system upgrade

Continuous work on the optics with a particular care on wiggler modelling resulted in a satisfactory agreement between model predictions and measurements of beta functions, dispersion, second order dispersion, chromaticities etc. [13]. The model reliably reproduces and predicts the machine optical functions allowing us to test very quickly for a wide range of lattice configurations (e.g. lattices with a momentum compaction factor varying from +0.04 to –0.036). The linear optics model has proven to be a good base for coupling correction, dynamic aperture optimization and nonlinear dynamics studies. At present DAΦNE operates with coupling corrected down 0.2-0.3% and a dynamic aperture as large as 14-15 $\sigma_x$.

The new interaction regions (doublet configuration) allowed a reduction of both IP beta functions without chromaticity increase together with a reduction of the non-linear forces due to the parasitic beam-beam crossings. Independent rotations of all IR quadrupoles provide local coupling correction and possibility of operation with arbitrary values of the detector solenoid field. For example, in the KLOE lattice configuration the values of $\beta_x$=1.6 m and $\beta_y$=1.7cm at the IP have been reached, to be compared to 4.5 m and 4.5 cm, respectively, of the initial design values. The horizontal emittance has also been reduced by about a factor 3 with respect to the design to further reduce the effect of the parasitic collisions. As a result we now collide routinely with 106-111 consecutive bunches out of 120 available; the short gap is still needed for ion clearing in the e- ring.

The poles of the wiggler magnets have been modified by adding longitudinally and horizontally shimmed plates [14]. Moreover, an extra sextupole component has been added on one of the terminal poles of each wiggler to make dynamic aperture optimization easier. Field measurements and beam tests showed a significant reduction of the second and fourth order terms in the fields and showed almost a factor 2 improvement in the energy acceptance. These modifications were essential for keeping a good lifetime despite the emittance reduction.

The longitudinal feedback systems were originally designed to damp only dipole multibunch oscillations excited by the beam interaction with parasitic high order modes. However, after filter modifications and overall tuning now they are capable to damp also the 0-mode and quadrupole instabilities [15]. In turn, with additional power amplifiers the transverse feedbacks can keep under control the transverse multibunch instabilities with a rise time as fast as 10 μs [16].

Installation of three octupoles in each ring [17] helped to increase the dynamic aperture and to compensate the lattice cubic nonlinearity that is a necessary condition to obtain good luminosity performance [18]. They proved to be particularly useful to compensate the effect of the nonlinearities from the parasitic beam-beam crossings, increasing the lifetime by about 30% when colliding at the maximum currents.

The wiggler ICE removal has reduced the electron ring beam coupling impedance by approximately a factor 2 [19]. Consequently several harmful effects such as more pronounced bunch lengthening of the electron bunches, lower microwave instability threshold, vertical beam size blow up above the instability threshold and bunch longitudinal quadrupole oscillations have been eliminated for the whole range of typical operating parameters. This has provided a 50% geometric luminosity increase during the last FINUDA run [12].

A dedicated work on background reduction including orbits and optical functions correction, working point fine tuning, sextupole and octupole strengths and scraper position optimization have allowed operating the collider in the "topping up" mode. This resulted in a significant increase of the integrated luminosity.

# CRAB WAIST SCHEME AT DAΦNE

Since several years the DAΦNE team has been discussing proposals and ideas aimed at the collider luminosity increase. Some of them are listed below:
- All wiggling machine to decrease the damping time
- Negative momentum compaction factor [20]
- PC compensation with current-carrying wires [21]
- Collisions under a very high crossing angle [22]
- Strong RF focusing [23]
- Crabbed waist collisions [9, 24].

Collisions in a negative momentum compaction lattice and PC compensation with the wires have been tested experimentally at DAΦNE while 3 different proposals for DAΦNE upgrade, DANAE [25], strong RF focusing collider [26] and crab waist upgrade [27], have been prepared in a written form.

After many discussions the DAΦNE team decided and the LNF Scientific Committee approved testing the crab waist idea at DAΦNE since:
- A high luminosity gain is expected
- Implementation of the crabbed waist scheme does not require many hardware modifications
- It fits the DAΦNE schedule: the modifications are being done during the planned shut down for the FINUDA detector roll-out and SIDDHARTA detector installation
- A moderate cost of the modifications
- If the test is successful the crabbed waist scheme can be adopted for the SuperB factory design [28].

Now a preparatory work for the crab waist scheme implementation at DAΦNE, including theoretical and numerical studies, hardware design, tests and installation, is well advanced.

## Crab Waist Concept

The Crabbed Waist scheme of beam-beam collisions can substantially increase collider luminosity since it combines several potentially advantageous ideas [24].

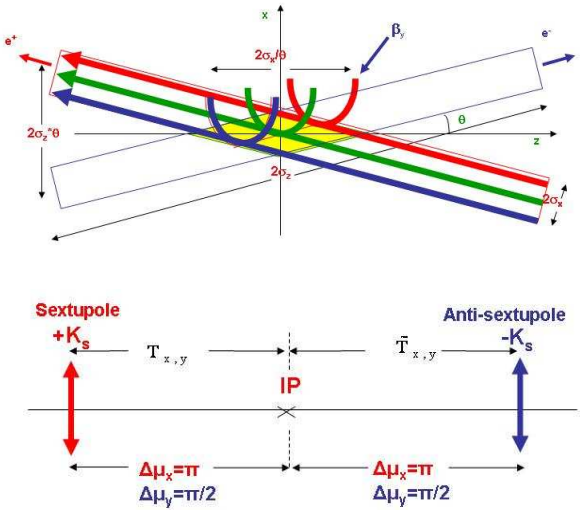

Fig. 3 Collision scheme with large Piwinski angle and crabbing sextupoles

The first one is large Piwinski angle. For collisions under a crossing angle $\theta$ the luminosity $L$ and the horizontal $\xi_x$ and vertical $\xi_y$ tune shifts scale as (see, for example, [29]):

$$L \propto \frac{N\xi_y}{\beta_y} \propto \frac{1}{\sqrt{\beta_y}}; \quad \xi_y \propto \frac{N\sqrt{\beta_y}}{\sigma_z \theta}; \quad \xi_x \propto \frac{N}{(\sigma_z \theta)^2}$$

Here the Piwinski angle is defined as:

$$\phi = \frac{\sigma_z}{\sigma_x} tg\left(\frac{\theta}{2}\right) \approx \frac{\sigma_z}{\sigma_x}\frac{\theta}{2}$$

with $N$ being the number of particles per bunch. Here we consider the case of flat beams, small horizontal crossing angle $\theta \ll 1$ and large Piwinski angle $\phi \gg 1$.

In the CW scheme described here, the Piwinski angle is increased by decreasing the horizontal beam size and increasing the crossing angle. In such a case, if it were possible to increase $N$ proportionally to $\sigma_z\theta$, the vertical tune shift $\xi_y$ would remain constant, while the luminosity would grow proportionally to $\sigma_z\theta$. Moreover, the horizontal tune shift $\xi_x$ drops like $1/\sigma_z\theta$. However, the most important effect is that the overlap area of the colliding bunches is reduced, as it is proportional to $\sigma_x/\theta$ (see Fig. 3). Then, the vertical beta function $\beta_y$ can be made comparable to the length of overlap area size (i.e. much smaller than the bunch length):

$$\beta_y \approx \frac{\sigma_x}{\theta} \ll \sigma_z$$

We get several advantages in this case:
- Small spot size at the IP, i.e. higher luminosity L.
- Reduction of the vertical tune shift $\xi_y$.
- Suppression of synchrobetatron resonances [30].

Besides, there are additional advantages in such a collision scheme: there is no need to decrease the bunch length to increase the luminosity as proposed in standard upgrade plans for B- and Φ-factories [31, 32, and 25]. This will certainly helps solving the problems of HOM heating, coherent synchrotron radiation of short bunches, excessive power consumption etc. Moreover, parasitic collisions (PC) become negligible since with higher crossing angle and smaller horizontal beam size the beam separation at the PC is large in terms of $\sigma_x$.

However, large Piwinski angle itself introduces new beam-beam resonances which may strongly limit the maximum achievable tune shifts (see [33], for example). At this point the crabbed waist transformation enters the game boosting the luminosity. This takes place mainly due to suppression of betatron (and synchrobetatron) resonances arising (in collisions without CW) through the vertical motion modulation by the horizontal oscillations [24]. The CW vertical beta function rotation is provided by sextupole magnets placed on both sides of the IP in phase with the IP in the horizontal plane and at $\pi/2$ in the vertical one (see Fig. 3). A numerical example of the resonance suppression is shown in Fig. 4.

*Beam-Beam Studies*

Beam-beam studies have verified the validity of the CW idea [24]. The effect of the betatron resonances suppression by the CW becomes evident when looking at the luminosity scan versus betatron tunes. Fig. 4 shows a luminosity scan in the tunes plane performed for DAΦNE in the SIDDHARTA configuration [27]. The scan on the left is with the CW sextupoles, the right one without. The red color corresponds to the maximum luminosity, the blue one to the minimum. With the CW many X-Y betatron resonances disappear or become much weaker, so the good working area is significantly enlarged, and the maximum luminosity is increased: a peak of $2.97 \times 10^{33}$ cm$^{-2}$ s$^{-1}$ compares with $L_{max} = 1.74 \times 10^{33}$ cm$^{-2}$ s$^{-1}$ without CW. Moreover in the CW collision a high luminosity can be obtained at the working points presently used at DAΦNE, like (0.09, 0.16). It should be noted that the worst luminosity value obtained in the CW collisions, $2.52 \times 10^{32}$ cm$^{-2}$s$^{-1}$, is still higher than the present best luminosity obtained at DAΦNE.

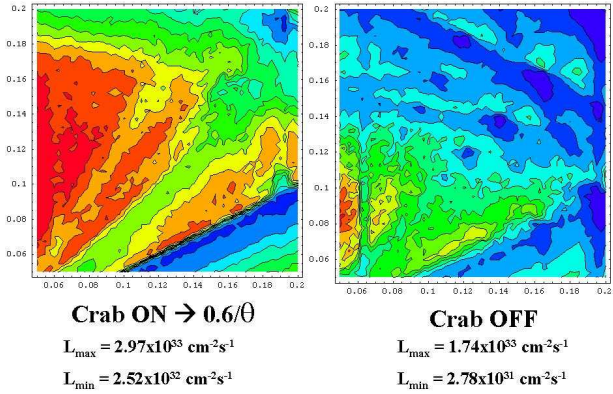

Crab ON → 0.6/θ
$L_{max} = 2.97 \times 10^{33}$ cm$^{-2}$s$^{-1}$
$L_{min} = 2.52 \times 10^{32}$ cm$^{-2}$s$^{-1}$

Crab OFF
$L_{max} = 1.74 \times 10^{33}$ cm$^{-2}$s$^{-1}$
$L_{min} = 2.78 \times 10^{31}$ cm$^{-2}$s$^{-1}$

Figure 4. Luminosity vs tunes scan.

An important limitation arising from the beam-beam interaction is the lifetime reduction. The beam-beam collisions create non-Gaussian tails in the transverse beam charge distributions. If the tails reach the physical or dynamic aperture the particles get lost, leading to lifetime degradation. In order to simulate the beam-beam induced tails the numerical code LIFETRAC [34] has been used. The CW sextupoles have been inserted in an implicit way, as lattice elements satisfying the CW conditions, i.e. having the required strength and betatron phase advances. Fig. 5 shows the beam distribution contour plots in the space of the normalized transverse amplitudes $A_x/\sigma_x$ and $A_y/\sigma_y$. For all the plots the maximum horizontal amplitude $A_x$ is $12\sigma_x$ and the vertical one $160\sigma_y$. The successive contour levels are at a constant ratio of $e^{1/2}$ between each other. Each column contains plots for different strengths of the crabbing sextupoles K: K = 1 means the exact crabbed waist condition (see [24]), for K = 0 the crabbing sextupoles are off.

A peak luminosity of about $3 \times 10^{33}$ cm$^{-2}$ s$^{-1}$ is achieved. The maximum luminosity is obtained for slightly lower sextupole strengths (K=0.6÷0.8) than required for the "exact" CW condition (K=1). The luminosity optimum corresponds also to the shortest distribution tails. With stronger or weaker sextupoles the tails start growing indicating possible lifetime problems. It is worth remarking that even without crabbing sextupoles (see the plots with K=0), a peak luminosity higher than $1.0 \times 10^{33}$ cm$^{-2}$ s$^{-1}$ can be achieved. Clearly the tails are much longer in this case. However, the lifetime can be improved with dynamic aperture optimization or by using slightly lower bunch currents.

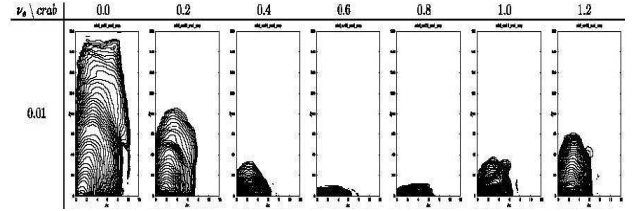

Figure 5. Distribution tails vs crabbing sextupole intensity

*Background Studies*

Machine backgrounds and lifetime will be dominated by the single Touschek scattering, as it is for the DAΦNE present configuration. Simulations of the Touschek effect with the CW scheme have been performed [27]. Particle losses due to Touschek effect are expected to be quite high with the Siddharta optics, mainly due to the smaller emittance. However, the longitudinal position of collimators has been optimized for the new optics and they are expected to be very efficient, even if a good compromise between losses and lifetime has necessarily to be found experimentally. In addition, careful design of the detector shielding is under way.

*Dynamic Aperture Studies*

Dynamic aperture (DA) studies have been performed with the Acceleraticum code developed at BINP [35], where a numerical algorithm is used to choose "the best" pairs of sextupole magnets in order to optimize the chromaticity correction. Moreover a tune working point was chosen which satisfies the requirements of high luminosity and large dynamic aperture. The "best pair" optimization method provides a dynamic aperture $\geq 20$ $\sigma_x$ off-coupling and $>250$ $\sigma_y$ full coupling, with an energy acceptance of ~1%.

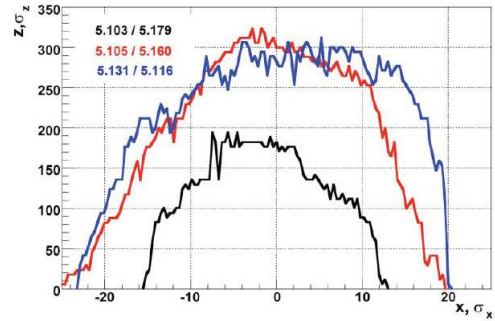

Figure 6. Optimized DA for different working points.

These values seem quite satisfactory to provide high luminosity and a successful experimental run. It is worth noting that one of the promising tune points {5.105, 5.16} practically coincides with the present operational values. Fig. 6 shows the optimized DA for three different working points, corresponding to good areas in the luminosity versus tunes plot.

*Hardware Modifications*

A layout of the upgraded DAΦNE is shown in Fig. 7, and the main hardware changes are briefly illustrated in the following. Details can be found in [36, 37].

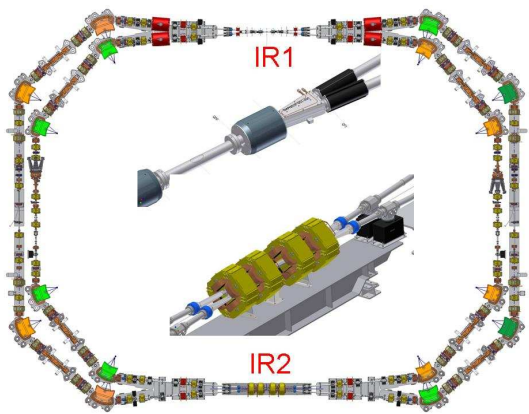

Figure 7. Upgraded DAΦNE layout.

The first interaction region has been modified [36] for the installation of the SIDDHARTA experiment and equipped with new quadrupoles designed to obtain lower $\beta^*$ at the IP. The total crossing angle has been increased from 30 mrad to 50 mrad. Existing sextupoles will be used as CW sextupoles. New beam pipes have been designed for this scheme. In the second interaction region the beams will travel through vertically separated vacuum chambers without low-β focusing.

Two new permanent magnet quadrupole doublets are needed in order to focus the beams to the smaller $\beta^*$ at the IP. The first quadrupole of the doublet is horizontally defocusing, common to both beams in the same vacuum chamber: it provides a strong separation of the beams. The following QFs are smaller, in order to fit separated beam pipes for the two beams.

New, fast kickers have been designed and built [38], based on a tapered strip with rectangular vacuum chamber cross section. The deflection is given by both the magnetic and the electric fields of a TEM wave traveling in the structure. Compared to the present DAΦNE injection kickers the new ones have a much shorter pulse (∼12 ns instead of ∼150 ns), better uniformity of the deflecting field, lower impedance and the possibility of higher injection rate (max 50 Hz).

Four new bellows [37] are placed in each sector, connecting pipes with circular cross section. The inner radius is 65 mm, the outer radius 80 mm and the length 50 mm. An RF shield is necessary to hide the chamber discontinuity to the beam. The coupling impedance of the structure has been evaluated in a frequency range from DC to 5 GHz and comes out to be very low.

Due to the new design solutions the total beam coupling impedance of the modified vacuum chamber is expected to be smaller with respect to the present one.

## CONCLUSIONS

DAΦNE has successfully completed its program for the three main physics experiments. Now a preparatory work is under way to implement a new crabbed waist collision scheme aimed at boosting the collider luminosity towards $10^{33}$ cm$^{-2}$s$^{-1}$. Commissioning of the modified DAΦNE is expected to start in November 2007.